\documentclass[sigconf]{acmart}

\AtBeginDocument{%
  }

\setcopyright{acmlicensed}
\copyrightyear{2018}
\acmYear{2018}
\acmDOI{XXXXXXX.XXXXXXX}
\acmConference[Conference acronym 'XX]{Make sure to enter the correct
  conference title from your rights confirmation email}{June 03--05,
  2018}{Woodstock, NY}

\acmISBN{978-1-4503-XXXX-X/2018/06} 
\usepackage[utf8]{inputenc}
\usepackage{booktabs}
\usepackage{graphicx}
\usepackage{subcaption}
\usepackage{caption}

\copyrightyear{2026}
\acmYear{2026}
\setcopyright{cc}
\setcctype{by}
\acmConference[CHI '26]{Proceedings of the 2026 CHI Conference on Human Factors in Computing Systems}{April 13--17, 2026}{Barcelona, Spain}
\acmBooktitle{Proceedings of the 2026 CHI Conference on Human Factors in Computing Systems (CHI '26), April 13--17, 2026, Barcelona, Spain}
\acmPrice{}
\acmDOI{10.1145/3772318.3791124}
\acmISBN{979-8-4007-2278-3/2026/04}

\begin{document}

\title[Rethinking Onboarding and Help-Seeking with Generative 3D Modeling Tools]{``\textit{I Just Need GPT to Refine My Prompts}'': Rethinking Onboarding and Help-Seeking with Generative 3D Modeling Tools} 

\author{Kanak Gautam}
\affiliation{%
  \institution{Computing Science, \\ Simon Fraser University}
  \city{Burnaby, British Columbia}
  \country{Canada}}
\email{kga74@sfu.ca}
\orcid{0009-0003-4831-6944}


\author{Poorvi Bhatia}
\affiliation{%
  \institution{Computing Science, \\ Simon Fraser University}
  \city{Burnaby, British Columbia}
  \country{Canada}}
\email{poorvi_bhatia@sfu.ca}
\orcid{0009-0003-8972-0492}

\author{Parmit K Chilana}
\affiliation{%
  \institution{Computing Science, \\ Simon Fraser University}
  \city{Burnaby, British Columbia}
  \country{Canada}}
\email{pchilana@sfu.ca}
\orcid{0009-0007-0173-1752}







\begin{abstract}
Learning to use feature-rich software is a persistent challenge, but generative AI tools promise to lower this barrier by replacing complex navigation with natural language prompts. We investigated how people approach prompt-based tools for 3D modeling in an observational study with 26 participants (14 casuals, 12 professionals). Consistent with earlier work, participants skipped tutorials and manuals, relying on trial and error. What differed in the generative AI context was where and how they sought support: the prompt box became the entry point for learning, collapsing onboarding into immediate action, while some casual users turned to external LLMs for prompts. Professionals used 3D expertise to refine iterations and critically evaluated outputs, often discarding models that did not meet their standards, whereas casual users settled for ``good enough.'' We contribute empirical insights into how generative AI reshapes help-seeking, highlighting new practices of onboarding, recursive AI-for-AI support, and shifting expertise in interpreting outputs.
\end{abstract}

\ccsdesc[500]{Human-centered computing~Human computer interaction (HCI)}

\keywords{Generative 3D Modeling, Help-Seeking, Software Learnability, Mental Models}

\maketitle

\section{Introduction}

Generative AI (Gen AI) tools promise to expand the reach of feature-rich design tools. By producing images, text, or 3D models from natural language prompts, they make it possible to start creating without first navigating the steep learning curves of feature-rich software. But insights from HCI remind us that lowering entry barriers often creates new forms of friction: unfamiliar affordances, opaque errors, and invisible labor such as troubleshooting, prompt iteration, or consulting external resources \cite{jintelligence13080103, karnatak2025expandinggenerativeaidesign, moore2023fAIlureNotes}. How, then, do users actually onboard these generative design systems, and where do they seek help when things go wrong?


We examine this question through the case of 3D modeling, a domain long known for steep learning curves and technical barriers. Advances in large multimodal models now allow users to produce 3D models from text-based prompts or sketches, promising faster workflows, richer exploration, and lower entry barriers \cite{li2024generativeaimeets3d, liu20233dalle, jun2023shapegeneratingconditional3d, lin2023magic3dhighresolutiontextto3dcontent}. These possibilities matter for two distinct groups of creators: \textit{professionals}, defined here as individuals with established expertise in 3D modeling through industrial practice, formal education, or freelance production; and \textit{casual makers}, who are individuals with little or no prior 3D experience but are motivated to explore 3D design and experiment with walk-up-and-use 3D printing services that allow first-time users to begin meaningful activity without prior preparation or training \cite{hudson2016understanding}. For professionals, generative systems can accelerate repetitive tasks, spark new directions in ideation, and complement production workflows \cite{liu20233dalle, woodruff2024howknowledge}. For casuals, such systems can open creative avenues that were previously difficult to learn, allowing them to bypass years of technical training and produce tangible 3D artifacts \cite{gaddam}.

\begin{figure*}[ht]
\centering
    \includegraphics[width=0.8\linewidth]{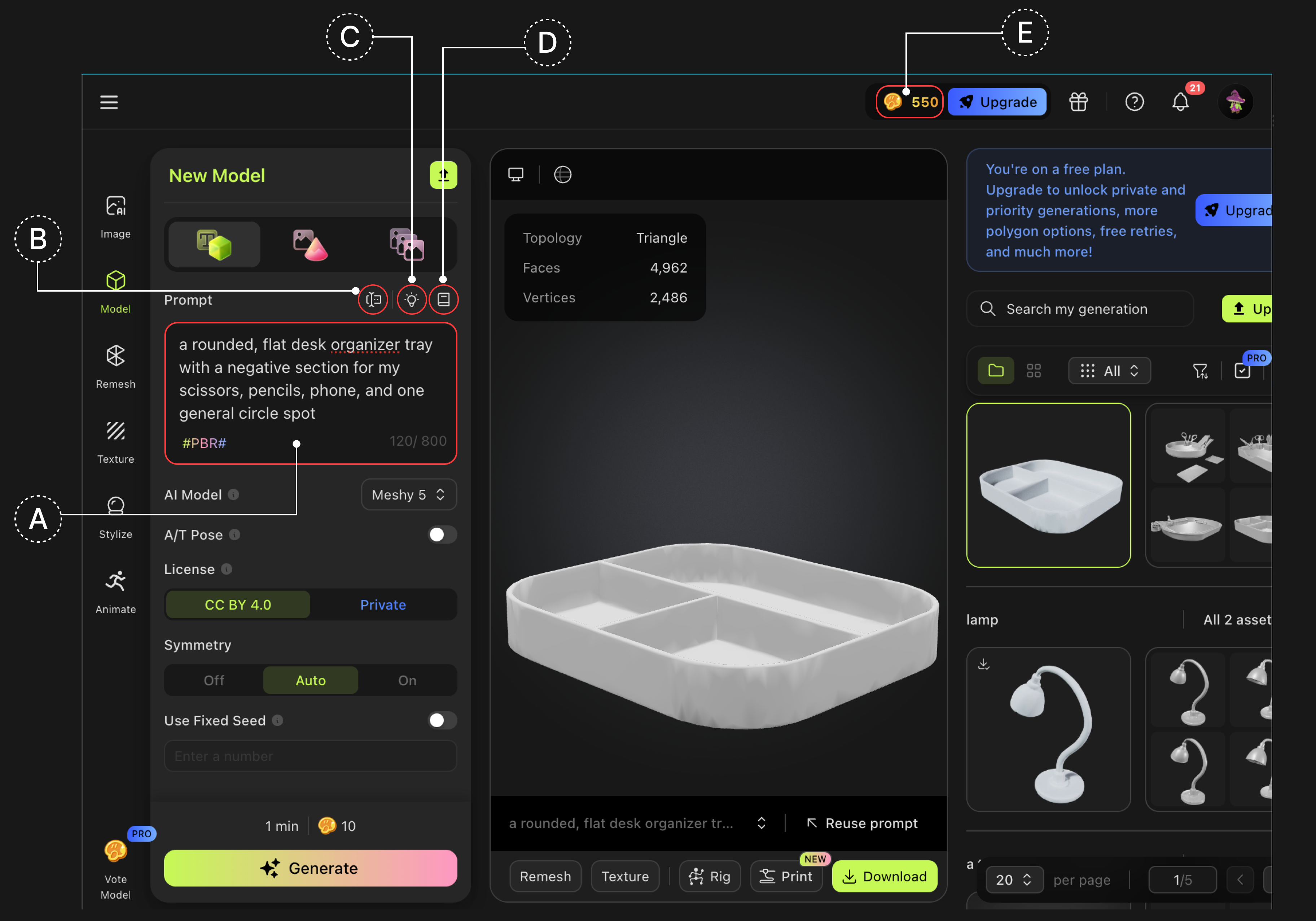}
\caption{Example of a Generative 3D modeling tool (\textit{Meshy AI}) used in the study. Its key features include: (A) Prompt input panel for describing the desired 3D model, including style and generation settings; (B) Built-in AI help option that refines vague prompts into detailed, structured prompts for more accurate generation; (C) Built-in library of example prompts providing ready-to-use templates for diverse modeling tasks; (D) Built-in tutorial link offering step-by-step guidance on using the tool effectively; (E) Credit and upgrade indicators showing remaining credits and subscription options. Spline AI interface was similar to this interface as well. Despite help and examples built into the UI, they were largely ignored by participants, many preferring to go to ChatGPT instead.}
\label{fig:meshy_interface}
\end{figure*}

However, despite this promise, there is limited understanding of how generative AI changes the actual work of onboarding and learning new software. In this paper, we use onboarding to refer to how users begin working with a new software or digital environment, how they learn its capabilities, explore its features, and establish an initial workflow \cite{PONI, useronboarding}. Prior HCI research has documented learnability barriers in complex software \cite{kiani2019beyond, ELLIOTT2002547, coyle2016learnability, Yun_2025}, help-seeking in online communities \cite{hou2024effectsgenerativeaicomputing, hou2025allroadsleadchatgpt}, and usability breakdowns in intelligent systems \cite{Khurana_2024}. Yet, these perspectives offer limited insight into how generative AI reshapes the cognitive and interactional work users must perform  \cite{casilli2024digitallaborinconspicuousproduction}. In creative, spatial workflows like 3D modeling, interactions are shaped by unpredictable outputs, ambiguous prompt affordances, and limited feedback when system interpretations fail. These challenges raise questions about how users, especially those with different levels of expertise navigate breakdowns, refine inputs, and decide when to seek support. Without attending to these dynamics, we risk overlooking the very frictions that could limit the broader participation in design that generative AI tools are meant to support.
Understanding how these frictions differ across  expertise levels is crucial for designing AI systems that truly augment, rather than complicate, creative work \cite{casilli2024digitallaborinconspicuousproduction, meluso2025invisiblelaboropensource}.
Motivated by these insights, our research questions were:

\begin{itemize}
\item {RQ1: How do casual and professional users form initial mental models when interacting with prompt-driven generative 3D modeling tools?}
\item {RQ2: How do users seek help and refine their workflows as they iterate designs with these generative 3D modeling tools?}
\end{itemize}

We conducted an in-depth observational study with 26 participants (14 casuals, 12 professionals) using state-of-the-art prompt-driven 3D modeling systems (\textit{Meshy} \cite{meshy_ai} and \textit{Spline} \cite{spline_ai2025}), where participants completed various modeling tasks using natural language. We also conducted semi-structured interviews to capture further insights into their experiences and decisions about how and when to seek support. Through task-based observations, think-aloud protocols, and interviews, we analyzed how participants navigated onboarding, formulated prompts, sought help, and refined their workflows over multiple iterations, with particular attention to differences between casual and professional users.

Our findings reveal new tensions in how generative AI lowers entry barriers for casual makers while simultaneously introducing frictions for professionals whose expectations for precision and workflow integration exceed current capabilities. Thematic analysis revealed three critical insights: both casual and professional users shared a ``prompt-first'' mental model, treating the text box as the primary entry point and bypassing tutorials or documentation; casual users frequently leveraged external AI tools such as ChatGPT to co-construct prompts, echoing concerns about prompt writing as a distinct practice of trial and error \cite{liang2025promptsprogramstoounderstanding, masson2022supercharging}.  Professionals often discarded AI-generated models as ``non-production-ready'', whereas casual users expressed satisfaction with outputs they perceived as sufficient for visualization or hobbyist 3D printing.

Our paper contributes empirical insights into how 26 casual and professional users onboard to generative 3D modeling tools. Our analysis reveals practices such as treating the prompt as a tutorial, using recursive ``AI-for-AI'' support, and engaging in resource-rational iteration shaped by credit-based systems. We also derive design implications for how generative AI can lower entry barriers for newcomers while supporting the precision and control required by professionals. By situating these insights within broader HCI work on human–AI collaboration, interpretability, and co-creativity, we extend understandings of onboarding and help-seeking in complex software. Our results suggest that in the era of Gen AI, software onboarding is emerging as a \textit{language-mediated experimentation} rather than traditional interface exploration \cite{carroll}. We highlight what is new: the emergence of AI-to-AI help-seeking and AI chaining as primary strategies for onboarding and troubleshooting. 
At the same time, our analysis shows that users’ underlying motivations remain consistent with long-standing patterns identified in prior HCI and learning work \cite{carroll, carroll1998minimalism, nelson, Aleven, Hutchins, kaptelinin2006acting}.
While our study focuses on 3D modeling, the observed dynamics likely extend to other domains where generative AI is  embedded into feature-rich creative tools.



\section{Related Work}

While our study highlights how generative AI reshapes onboarding and help-seeking, these issues are not new to HCI. Prior work has long examined how users learn feature-rich software, seek support through documentation and communities, and adapt to intelligent systems.  HCI has grounded these behaviors in theories of learnability, self-regulated learning, distributed cognition, and trust in automation \cite{carroll, nelson, kaptelinin2006acting}. What remains less understood is how generative AI changes these dynamics, collapsing tutorials into prompts, shifting help-seeking across tools, and reframing expertise. We therefore turn to related work on software learnability, help-seeking, and generative AI in creative practice.

\subsection{Learning and Help-Seeking in Complex Software}

Feature-rich creative tools, such as CAD systems, game engines, and 3D modeling platforms, have long been recognized for their steep learning curves. Prior HCI studies document the \textit{vocabulary problem} \cite{furnas1987thevocabularyproblem, hudson2016understanding, rieman1996afieldstudy}, where novices struggle to find help using correct terminology, and show that users often bypass manuals or tutorials \cite{Angela}. This pattern has been classically described as the paradox of the active user \cite{carroll}, which explains why people prioritize immediate task progress over systematic learning even when this slows long-term performance. Documentation, video tutorials, and community forums provide flexible but inconsistent support: they may be outdated, difficult to search, or misaligned with the user’s goals \cite{robillard2011afieldstudy, brandt2009twostudies, Zhang_2021}. Alongside such embedded supports, prior work has shown that trial-and-error remains a dominant learning strategy in complex software, where users iteratively test commands or features until achieving a desired outcome. Studies of CAD, programming environments, and creative tools consistently observe this incremental process as a practical, if sometimes inefficient, way to onboard and make progress \cite{masson2022supercharging, learnbytrial}. This illustrates how onboarding and help-seeking have historically been distributed across multiple pathways, each carrying friction. Researchers responded by embedding \textit{in-situ support systems} \cite{chilana2012lemonaid, edwards2004examplecentric, brandt2010examplecentric} such as interactive walkthroughs, examples, or context-aware agents directly into interfaces to scaffold learning within workflows. These systems aim to reduce friction by surfacing assistance within the user’s workflow, lowering barriers to experimentation and learning. However, these approaches often target deterministic software behaviors and may fall short when integrated into generative systems with stochastic outputs, highlighting the need for more adaptable support that flexibly responds to unpredictable model behavior. To the extent these approaches can be integrated into generative systems with stochastic outputs and  unpredictable model behavior is an open research question. Our work adds new insights about how onboarding and help-seeking unfold in generative AI–assisted 3D modeling, where users must not only learn tool interfaces but also grapple with interpreting unpredictable outputs. We show how both casual and professional users navigate these challenges by blending self-written prompts, external resources, and AI-based guidance, revealing new forms of friction and support-seeking strategies that extend beyond those documented in prior studies of deterministic software.

\subsection{Invisible Labor and Hidden Costs in Intelligent Systems}

Beyond software learning, recent work in HCI and software engineering is illuminating the \textit{invisible labor} that sustains interaction with complex and intelligent systems \cite{fox2023patchwork, casilli2024digitallaborinconspicuousproduction, meluso2025invisiblelaboropensource, heer_agency_2019, zamfirescu2023whyjohnny, IsItMeOrAI, oppenlaender2024promptingaiartinvestigation}. Studies of end-user programming and professional software use show that users often rely on trial-and-error loops, incremental debugging, and extensive online searches, forms of work that are rarely visible in final output \cite{kiani2019beyond, carroll, brandt2009twostudies}. With the advent of large language models, this hidden effort shifts toward \textit{prompt crafting and iterative refinement}. Research on AI-assisted coding demonstrates how novices may treat AI assistants as black-box tutors while experts integrate them selectively, both groups expending unseen cognitive and emotional labor to vet outputs, recover from errors, or adapt workflows. Broader human–AI collaboration studies underscore how unpredictability, lack of transparency, and uneven control exacerbate these hidden costs \cite{peng2023impactaideveloperproductivity, Zhang_2023, vaccaro_when_2024}. This body of work provides a critical lens: while generative AI appears to ``lower barriers,'' it may instead redistribute labor into subtler, less visible forms that shape trust, efficiency, and learning. Our study complements this body of work by examining invisible labor in the domain of generative 3D modeling. Unlike coding or end-user programming, where trial-and-error loops or debugging are already established practices, 3D modeling introduces distinct forms of hidden effort: crafting prompts that balance form and function, repeatedly iterating when outputs misalign with intent, and turning to external resources when system guidance falls short. By surfacing how both casual and professional users distribute their labor across prompting, external help-seeking, and post-processing, our findings highlight how generative systems may not simply reduce complexity but shift it into less visible forms of interaction. In doing so, we extend HCI conversations on invisible labor from software engineering contexts into creative design tools, showing how unpredictability and stochasticity uniquely reshape onboarding and learning in this domain.

\begin{table*}[ht]
\centering
\small  
\setlength{\tabcolsep}{3pt}  
\caption{Overview of participant demographics, including gender, age group, academic or professional background, and tool expertise of the professionals) across all 26 participants in the study. Participant IDs with a `C' indicate casual users, while IDs with a `P' indicate professional users.}
\label{tab:participants}
\begin{tabular}{|cccc|cccc|}
\hline
\textbf{ID} & \textbf{Age} & \textbf{Background} & \textbf{3D Tool Expertise} &
\textbf{ID} & \textbf{Age} & \textbf{Background} & \textbf{3D Tool Expertise} \\
\hline
C01 & 19--24 (M) & Computer Science & None & C10 & 19--24 (F) & Health Sciences & None \\
C02 & 19--24 (F) & Designer & None & C11 & 19--24 (F) & Computer Science & None \\
C03 & 19--24 (F) & International Studies & None & C12 & 25--34 (F) & Web Design & None \\
C04 & 19--24 (F) & Statistics & None & C13 & 55--64 (F) & Physics & None \\
C05 & 19--24 (F) & Arts & None & P14 & 19--24 (F) & Computer Science & AutoCAD \\
P06 & 25--34 (M) & Game Design & Blender, Maya & C15 & 25--34 (M) & Computer Science & None \\
P07 & 25--34 (M) & 3D Printing & AutoCAD & P16 & 25--34 (F) & Interactive Arts \& Tech. & AutoCAD, Rhino \\
C08 & 19--24 (F) & Computer Science & None & P17 & 35--44 (M) & Game Design & Blender, Maya \\
P09 & 25--34 (F) & 3D Printing & AutoCAD, Revit & P18 & 35--44 (M) & AR/VR & 3Ds Max \\
P19 & 19--24 (GQ) & Computer Science & Unity, Maya & C20 & 25--34 (F) & Computer Science & None \\
P21 & 35--44 (M) & Computer Science & Fusion360, AutoCAD & C22 & 19--24 (F) & Health Science & None \\
P23 & 19--24 (M) & Electrical Eng. & AutoCAD & C24 & 25--34 (M) & Computer Science & None \\
P25 & 19--24 (F) & Computer Science & Fusion360 & P26 & 19--24 (M) & Computer Science & Fusion360 \\
\hline
\end{tabular}
\end{table*}

\subsection{Generative AI and Prompt-Based Creative Practice}

In addition to software development, recent advances in generative AI have opened new possibilities in creative domains such as art, text, music, and 3D modeling \cite{bai2024progressprospects3dgenerative, epstein_art_2023}. Prompt-based systems like DALL·E and Firefly illustrate how natural language interfaces collapse low-level commands into expressive requests, while research has begun examining prompting as a skill that shapes quality, diversity, and control of outputs \cite{ramesh2021zeroshottexttoimagegeneration, adobe2025firefly, long2025makesgoodnaturallanguage}. In 3D, models such as DreamFusion, Magic3D, and Shap-E demonstrate pipelines that translate text or images into meshes, while sketch-to-3D systems more closely align with traditional design workflows \cite{poole2022dreamfusiontextto3dusing2d, lin2023magic3dhighresolutiontextto3dcontent, jun2023shapegeneratingconditional3d, chen2023deep3dsketch3dmodelingfreehand}. These innovations have given rise to casual-facing tools (e.g., Meshy AI, Spline AI) that echo ``walk-up-and-use'' fabrication paradigms, as well as professional use cases where AI outputs are exported for refinement in Blender or Unity \cite{buechley2010lilypad, blenderOrg, unityTechnologies2025}. Yet HCI scholarship remains limited in understanding \textit{how real users actually adapt}: casual users may treat AI as a shortcut to producing printable forms, while professionals critique models against production standards and discard outputs deemed unusable. This highlights a gap: while prior work studies prompting and creative AI broadly, little is known about how \textit{onboarding and help-seeking} play out when generative AI becomes the entry point to complex domains like 3D modeling. Our study addresses this gap by examining how both casual and professional users onboard generative AI–assisted 3D modeling tools and seek help when encountering uncertainty. We show how participants engage in trial-and-error loops, and turn to external resources when system support falls short. By surfacing these practices, we extend HCI scholarship on prompt-based creative practice beyond 2D media and coding, situating it within 3D workflows where precision, production constraints, and usability demands create unique forms of friction. In doing so, we highlight how generative AI does not simply collapse complexity but redistributes it into new learning and labor practices.

\section{Method}

The goal of this research is to explore how users with varying levels of expertise in 3D design, ranging from casual makers with little to no prior experience to professional designers with formal training onboard, seek help when navigating generative AI modeling tools for the first time. We take an interpretivist stance \cite{Braun01012006}, aiming to surface participants’ meaning-making and invisible labor, rather than to measure performance outcomes. We used a qualitative approach as it is well-suited for exploring how practices of onboarding and help-seeking unfold in situated, real-world use of generative tools. We started with an observational task-based study using two popular generative 3D modeling tools and followed-up with in-depth interviews to capture users’ reflective understanding of the challenges they faced, the strategies they adopted, and how they would prefer such systems to support creative practices and learning processes. 

\begin{figure*}[t!]
    \centering
    \begin{subfigure}[t]{0.4\textwidth}
        \centering
        \includegraphics[height=1.5in]{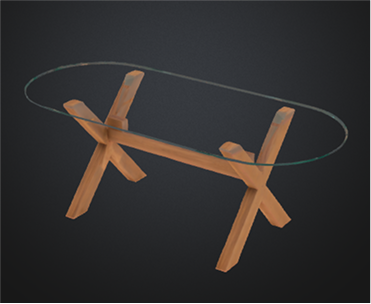}
        \caption{Target model used in Task 1}
        \label{fig:task1_target_img}
    \end{subfigure}%
    ~ 
    \begin{subfigure}[t]{0.4\textwidth}
        \centering
        \includegraphics[height=1.75in]{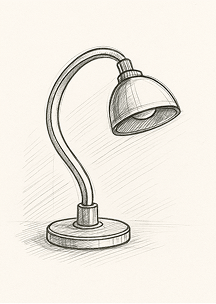}
        \caption{Target model used in Task 2}
        \label{fig:task2_target_img}
    \end{subfigure}
    \caption{Target reference models provided to participants for modeling tasks. (a) Task 1: A glass-top wooden table (adapted from \cite{kiani2019beyond}); (b) Task 2: A desk lamp with an adjustable arm. These tasks were designed consulting experts in 3D modeling and were designed in target generative AI tool (Meshy or Spline).}
    \label{fig:target_imgs}
\end{figure*}

\subsection{Participant Recruitment and Grouping}

We recruited a total of 26 participants (10M, 15F, 1 Gender Queer), as shown in Table~\ref{tab:participants} and divided them into two categories:
\begin{itemize}
    \item \textit{Casual Users (C):} Participants who had little or no formal training in 3D modeling and were novices in this space \cite{hudson2016understanding, jahanlou2024howexamplebased}. This group included students from both technical and non-technical backgrounds, hobbyists interested in creative tools, and other users curious about generative 3D design. None of them had used CAD tools extensively and were not familiar with professional modeling workflows. (n = 14)
    \item \textit{Professional Users (P):} Participants who had formal training and/or industry experience in 3D modeling, such as CAD designers, architects, industrial designers, game designers and 3D artists. All had previously used tools such as Fusion 360, Rhino, Blender, or SolidWorks in professional settings and were familiar with modeling constraints for production and 3D printing. (n = 12)
\end{itemize}

Casual users were recruited through university mailing lists and flyer advertisements, while professionals were recruited through professional design networks, local makerspaces, and snowball sampling. Our goal was to cover a broad spectrum of experience levels in 3D modeling, ranging from absolute novices to trained professionals. We ensured a diverse sample across domains, educational backgrounds, and age groups. Table~\ref{tab:participants} summarizes the demographics, backgrounds, and experience levels of our participants. Most of our participants never used any Generative 3D modeling application before. 

As a research team with backgrounds in HCI and experience in both professional 3D design and novice creative tool use, we recognize our positionality shaped what we attended to in data collection and interpretation. We practiced reflexivity through regular debriefs, challenging our assumptions about expertise, creativity, and tool usability.

\subsection{Study Context and Tool Selection}


For our study, we sought tools that would allow users to generate 3D assets using natural language prompts, reference images, or texture specifications. After exploring available platforms and consulting with 3D design experts, we selected two widely used web-based generative 3D modeling systems, Meshy AI \cite{meshy_ai} and Spline AI \cite{spline_ai2025}, that represent the current wave of commercially deployed AI-powered modeling tools. 

A key characteristic of these systems is that they operate on a credit, or usage-based pricing model: each model generation, refinement, or upscaling action consumes a predefined number of credits. This pattern is not unique to 3D modeling but increasingly common across the broader GenAI ecosystem. Major creative-AI platforms such as Runway (video generation) \cite{runway2024gen2}, Adobe Firefly (image and design generation) \cite{adobe2025firefly}, and Sora \cite{openai2024sora} also meter usage through generative credits; similarly, multimodal APIs from OpenAI \cite{openai2024}, Stability AI \cite{stabilityai2024} and Google rely \cite{googledeepmind2024} on token- or generation-based billing rather than unlimited access. These examples reflect a broader industry shift toward usage-based pricing for AI-enhanced creative work, particularly in feature-rich domains where computational cost varies by output quality and model complexity. Although our study did not explicitly aim to investigate credit behavior, credit constraints organically emerged as a strong influence on user decision-making during pilot testing. Hence, to maintain the ecological validity of the study, we decided to test participants under credit-based situations.

\subsection{Task Design and Rationale}

To investigate how users approach and learn generative 3D modeling, we designed three tasks reflecting common entry points in creative and professional workflows. These tasks elicited behaviors ranging from initial prompt instincts to visual input interpretation and open-ended ideation. We refined them through pilot testing and consultation with practicing 3D designers.

\textit{\\Task 1: Text-to-3D Generation}

The first task was to generate a 3D model using the tool's natural language prompt interface. This task examined users' initial mental models when approaching generative 3D modeling--whether they described high-level features (e.g., overall shape or intended function) or low-level ones (e.g., specific parts, materials, or fine details of the model). We also observed strategies for handling uncertainty, such as  trial and error and use of help within the application or through external resources (participants were encouraged to use any help resource of their choice for all tasks). 

The target object was a glass-top table with a wooden base, as shown in Figure~\ref{fig:task1_target_img}, adapted from Kiani et al.'s \cite{kiani2019beyond} study. We selected this as it has a moderately complex structure involving multiple materials and surface expectations, useful for assessing how participants translate visual ideas into text prompts. The reference model was created in advance by the researchers using the same AI tools (Meshy and Spline) and refined after consulting professionals in this domain.

This task surfaced how participants (especially those unfamiliar with the 3D modeling vocabulary) navigate the prompt formulation, feature interpretation, and model inspection when no explicit visual cues are provided.

\textit{\\Task 2: Image-to-3D Generation: Sketch-Based Visual Modeling}

In Task 2, participants generated a 3D model from a sketch using the generative 3D tool's  image-to-3D feature. This task was inspired by early findings from our pilot study, where multiple professional participants expressed a preference for visual input as a more natural part of their creative process, as many workflows begin with sketches, moodboards, or reference images rather than detailed text prompts.

We provided participants with a minimal line drawing of a small side table lamp, as shown in Figure~\ref{fig:task2_target_img}, and asked to generate a 3D model that aligns with the sketch. This allowed us to assess how participants interpret and control the fidelity of the AI output against a visual target, and how they perceive tool capability when working with concrete visual constraints. 

This task revealed differences between casual and professional participants in leveraging visual scaffolds and evaluating the quality of 2D to 3D translation across generative platforms.

\textit{\\Task 3: Open-Ended Generation: Creative or Goal-Oriented Design}

The third task was tailored by the participant group. Casual users were asked to design a simple, fun, or useful object for 3D printing, revealing their motivations, ideation strategies, and perceptions of printability and tool constraints. Professional users were asked to create a basic model for potential professional use (e.g., a prototype enclosure), allowing us to examine how they assessed mesh quality, editability, and readiness for downstream CAD or production tools.

Overall, this task surfaced where participants saw value in AI assistance, whether for ideation, sketching, rapid prototyping, or final production, and how expectations shifted with expertise.



\subsection{Study Procedure}

Each study session was conducted in person in the lab or remotely over Zoom and lasted between 60 and 75 minutes. All remote participants joined from their personal computers in their own environments to preserve ecological validity and simulate natural use conditions. The procedure was structured into three phases: orientation, task execution with concurrent think-aloud, and a post-task semi-structured interview.

\subsubsection{Orientation and Setup (5–10 min)} 
Participants first completed a pre-study questionnaire designed to capture demographics, prior experience with 3D modeling tools, and familiarity with generative 3D design. We next randomly assigned participants to either the Meshy AI or the Spline tool condition using researcher-provided accounts. The researcher running the study provided a brief orientation to the platform interface, covering basic navigation and functional features. Importantly, no task-specific examples or strategies were shared; this decision was intentional to recreate the self-directed onboarding conditions under which users typically encounter generative AI tools. Participants were informed that they could freely use any help built into the tool and any external resource they may normally rely on (e.g., Google, YouTube tutorials, platform documentation, or ChatGPT) but that the researcher would not provide direct task guidance.

\subsubsection{Task Execution and Think-Aloud (35–45 min).} Participants completed the three modeling tasks in a fixed order: text-to-3D generation, image-to-3D generation, and the open-ended design task. Tasks were presented in a fixed order to scaffold participants from lower- to higher-constraint activities, reflecting how generative tools are typically encountered. While we considered counterbalancing, we prioritized a consistent progression to ease learning for novices and to keep the focus on interaction patterns rather than comparative performance.

To explore how credit systems shape behavior, we introduced two contrasting contexts. Half of the participants (n=13) were \textit{credit-aware}: they were told about the platform’s credit-based generation model and given just enough credits to attempt each task twice. This setting foregrounded how resource limits influenced strategies such as cautious iteration, prompt abandonment, or seeking external assistance. The remaining participants (n=13) were \textit{credit-unaware}: they received accounts with unrestricted credits and no mention of the system, reflecting the experience of encountering such tools without explicit awareness of usage constraints. This contrast was intended to surface potentially diverse strategies and perceptions, not to produce comparative performance metrics.


During task execution, all participants were instructed to think aloud, verbalizing their reasoning, frustrations, and decision-making in real time, including the remote participants who additionally shared their screens. 
The researcher captured prompt formulations, credit use (or absence of credit concerns), tool navigation patterns, and any external help-seeking behavior through screen recordings. The researcher remained a silent observer, intervening only in the case of technical difficulties, to preserve the authenticity of participants’ self-directed learning and troubleshooting. 

After completing each task, participants filled out a brief 7-point Likert scale questionnaire to rate their perceptions of the platform for that task, their satisfaction with the generated outputs, the predictability and trustworthiness of the system’s behavior, and their perceived likelihood of using such a workflow in the future. Embedding the questionnaire after each task allowed us to capture participants’ impressions while they were still fresh, avoiding retrospective bias and enabling direct comparison of attitudes across tasks.

\subsubsection{Post-Task Semi-Structured Interview (15–20 minutes).} We briefly interviewed each participant to probe into participants’ expectations of generative 3D modeling tools, the troubleshooting strategies they employed for prompt iteration and error correction, and their evaluation of the usefulness and trustworthiness of AI outputs. We also asked participants to situate these tools in relation to their broader creative or professional practices. For professional participants, we further explored perceptions of model quality and compatibility with downstream workflows (e.g., CAD tools, 3D printing pipelines). The interview concluded with open-ended reflections on participants’ ideal form of AI assistance and the barriers they perceived to adoption.

This structured procedure allowed us to capture both in-the-moment behaviors during interaction and post-hoc reflections on onboarding, help-seeking, and workflow integration \cite{joanna}.

\subsection{Help Resources and Interaction Environment}

To preserve ecological validity, participants were explicitly encouraged to use any resources they would normally draw upon in real-world tool exploration. Rather than restricting access to predefined materials, we sought to observe how participants engaged in self-directed help-seeking when confronted with uncertainty or system limitations \cite{kiani2019beyond}. As such, participants were free to consult the platforms’ built-in tutorials or tool-tips (as shown in Figure~\ref{fig:meshy_interface}), external video tutorials and online documentation, community forums, or general-purpose search engines such as Google. We also allowed participants to make use of external AI tools (e.g., ChatGPT) during the study.

Credit consumption was tracked through platform logs, and participant reactions to credit depletion or credit awareness were closely noted. These reactions often surfaced spontaneously in the course of interaction, particularly among casual users, and shaped the depth of their exploration, their reliance on external help resources, and their overall satisfaction with the generated models. This approach allowed us to capture credit-related concerns as they naturally emerged, rather than foregrounding them as an explicit study variable.

\subsection{Data Collection and Recording}

We employed a multimodal data collection strategy to capture both participants’ direct interactions with the tools and their reflective accounts of the experience. Each session was recorded in full, including participants’ screens, audio, and video feeds, with prior consent. The screen recordings documented fine-grained interaction details such as prompt edits, help seeking approaches, and inspection of generated models, while the audio channel preserved concurrent think-aloud commentary as well as the post-task interviews.

In addition to recordings, the first author took notes during the sessions to document contextual observations, for example, moments of hesitation, nonverbal cues, or spontaneous remarks that might not be fully captured in transcripts. Where participants used ChatGPT to refine or troubleshoot their prompts, we archived those conversational exchanges to trace instances of cross-tool chaining. Platform-level logs were also collected when available, providing records of input prompts, generated outputs, and credit consumption.

All audio recordings were transcribed and combined with the screen and system logs to provide a comprehensive dataset of user behavior. The observer notes and ChatGPT logs offered additional contextual layers. Together, these complementary sources allowed us to triangulate across interactional, system-level, and self-reported data, strengthening credibility by comparing what participants said, what they did, and how the system responded. This triangulation reflects an interpretivist orientation, treating multiple perspectives as complementary in constructing meaning rather than as checks on a single truth \cite{Braun01012006}. 

\subsection{Data Analysis}

We analyzed the data using reflexive thematic analysis following Braun and Clarke’s \cite{Braun01012006} six-phase method to identify patterns in how participants interacted with generative 3D modeling tools and how these patterns varied across expertise levels and credit conditions. 

Two of the authors collaboratively analyzed all transcripts, screen recordings, and field notes using a reflexive thematic analysis approach. We iteratively refined the codebook through repeated discussion, balancing descriptive accounts of user actions with interpretive insights into participants’ reasoning, learning strategies, and expectations. To establish shared understanding of code boundaries early in the process, both researchers independently coded an initial calibration subset of the data. For this subset, we calculated Cohen’s kappa (0.80) as a pragmatic check on alignment in code application, not to claim objectivity or coder neutrality. Disagreements were discussed reflexively and used to deepen interpretation and refine the analytic lens rather than treated as reliability failures. The remaining data were then coded through iterative, collaborative analysis consistent with reflexive thematic analysis practices.

Codes were then clustered into broader themes such as \textit{AI chaining} (e.g., using ChatGPT to refine prompts), \textit{trial-and-error adaptation} and its constraints, perceptions of model quality and production readiness, expertise-driven workflow differences, trust breakdowns, and \textit{help-seeking avoidance} in favor of self-guided exploration. Light quantification, such as counting participants who used ChatGPT or avoided help resources, was used to contextualize the prevalence of these behaviors within our qualitative themes.

\subsection{Ethical Considerations and Compensation}
The study was approved by the Institutional Research Ethics Board. All participants were informed of their right to withdraw at any point. Personally identifiable data was removed from all recordings and transcripts. Participants received a \$15 gift card as a token of appreciation for their time.

\section{Results}

\subsection{Initial Mental Models: Prompting vs. Seeking Guidance}


All participants (14C, 12P) began the first task by opening their assigned generative 3D modeling tool. Their strategies diverged immediately: some entered a self-formulated description directly into the prompt field, while others consulted external resources, such as ChatGPT, Google Search, or the tool’s documentation, before providing their first input. These early choices reflected participants’ assumptions about how the system should be used to generate output.

\begin{table*}[!ht]
\centering
\caption{Variations in prompting styles between professional and casual participants. Professional prompts were generally more structured, precise, and detailed, often specifying dimensions, materials, and design constraints, whereas casual prompts tended to be simpler, less specific, and conversational in tone. The average word length for a professional prompt was 83 vs for casuals it was 35.}
\label{tab:prompts_table}
\begin{tabular}{p{0.48\textwidth} p{0.48\textwidth}}
\toprule
\textbf{{Professional Prompts}} & \textbf{{Casual Prompts}}\\
\midrule
\textbf{P21:} \textit{Generate a glass coffee table with the legs made out of wood. The legs are Xs where the bottom of the X are the four feet of the table, and the top supports the glass of the table. The glass of the table is a rectangle with the corners rounded enough to be a full circle. The cross brace is a single piece with roughly 1x4x20 dimensions and is in the center of the X of the leg braces}. & \textbf{C04:} \textit{a table with elliptical glass top and wooden legs in the shape of an X} \\
\midrule
\textbf{P23:} \textit{suppose you are building a table with glass top with it having rounded corners with some Radius have crossed wooden legs at both sides, make it simple with glossy top and ensure that these legs are connecting with each other} & \textbf{C05:} \textit{can you help me with an image of coffee table- which consists of wooden legs, and a glass top. specifically, there should be 4 wooden legs and a semi circled glass top.} \\
\midrule
\textbf{P16:} \textit{I want to generate a table in which the table top material is glass, both semi-transparent and the legs are two parallel cross shapes with an element connecting them. these are all light coloured wood with square section} & \textbf{C10:} \textit{image of a coffee Table with wooden frame and glass top} \\
\bottomrule
\end{tabular}
\end{table*}

\begin{figure*}[!h]
    \centering
    \includegraphics[width=0.95\linewidth]{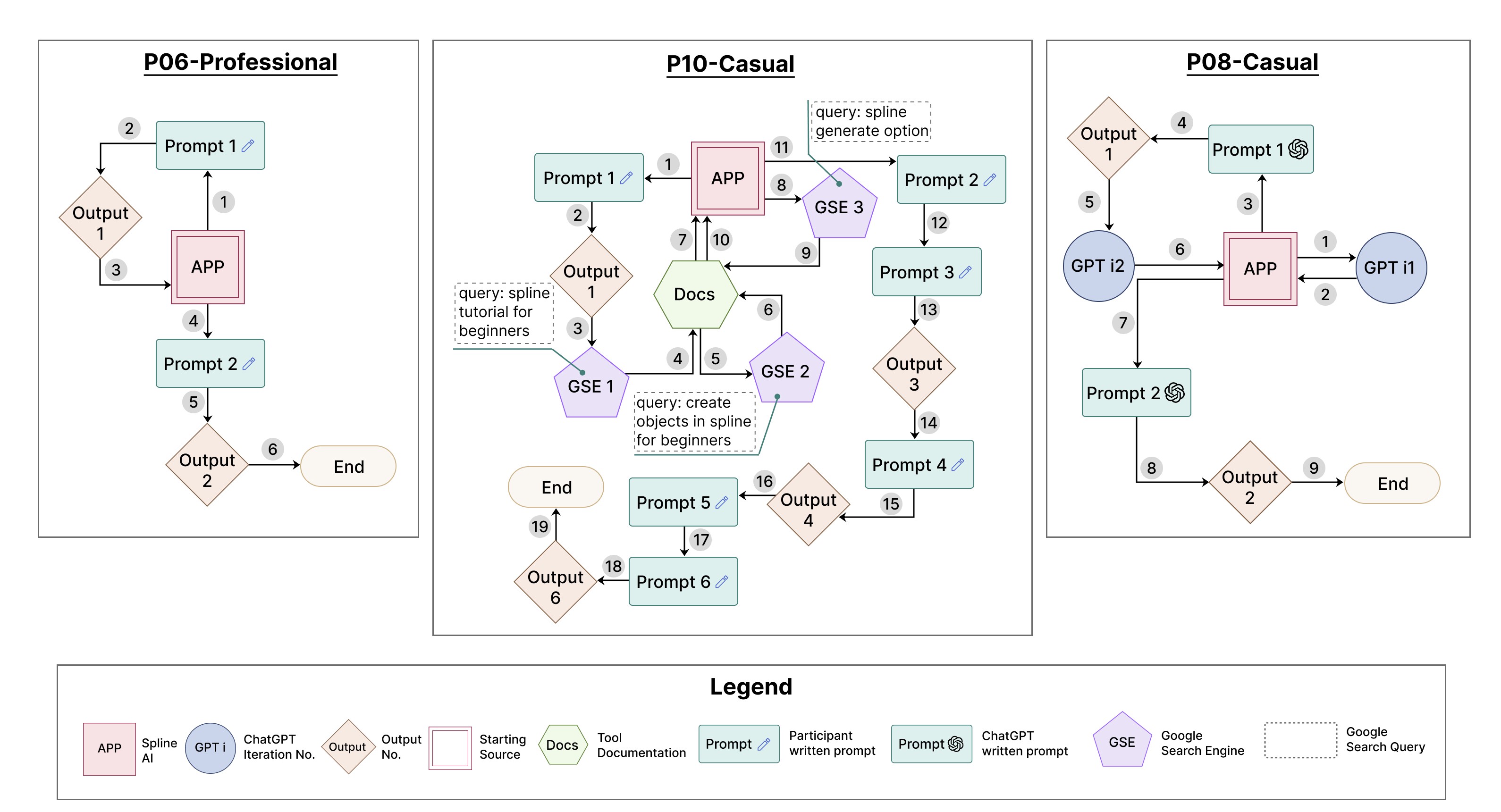}
    \caption{Workflow visualizations of three participants illustrating the distinct approaches to onboarding and iterating with the generative 3D modeling tool. \textbf{Left (P06–Professional):} Linear path where the the user did not take any external or in-built help. \textbf{Center (P10–Casual):} The casual user cycled across multiple help-seeking resources, including tool documentation (docs), Google search engines (GSE), while iteratively generating prompts and evaluating outputs. This pathway demonstrates a ``trial-and-error'' orientation, with heavy reliance on external aids to understand of both prompting and tool capabilities. \textbf{Right (P08–Casual):} An AI-chaining approach where the participant depended on ChatGPT to draft and refine prompts before using them in the generative application, highlighting how casual users offloaded the burden of prompt crafting to external AI. Together, these three trajectories capture the major approaches in our study: (1) Self Formulated prompts, (2) Exploration scaffolded by external documentation and search, and (3) AI-for-AI prompting, where users relied on generative models to learn how to interact with other generative systems.}
    \label{fig:placeholder}
\end{figure*}

\subsubsection{Beginning with Self-Formulated Prompts as the Default Strategy}
Even though none of our participants had used generative 3D tools before, most participants (18/26) began their first task by entering their own prompt directly into the tool. This strategy was observed among half of casual users (7C/14C) and almost all professionals (11P/12P). Rather than consulting an external resource or built-in tips, participants moved directly to the blank text box to compose a descriptive input prompt. Their behavior suggests that the text box was interpreted as an immediate \textit{call to action}, and that these participants felt confident starting without additional guidance.

This behavior contrasts with prior studies (e.g., \cite{kiani2019beyond}) that showed that casual users often first turned to Google search and YouTube videos when learning an unfamiliar 3D modeling tool. In our study, however, none of the casual participants sought out YouTube or other video-based tutorials before generating their first prompt. One participant explicitly articulated this shift in mindset compared to how he would learn other types of software, stating, \textit{``Well, AI tool is made for everyone, and especially for non-professional use, so… they usually tend to be beginner-friendly, so I don't need any tutorials''} (C15). 

A small subset (4C) consulted the official documentation, mainly to familiarize themselves with the interface. However, they quickly abandoned it (within a few seconds) when it failed to provide actionable guidance.


Among participants who began by entering prompts directly into the interface, casual and professional users showed stark differences in specificity. Professionals drew on prior CAD and 3D modeling experience to produce detailed, constraint-oriented prompts, often including dimensions, structural references, or geometry specifications (see Table~\ref{tab:prompts_table}). Casual participants, by contrast, relied on shorter, less precise descriptions, or phrased conversationally. An analysis of word counts of prompts across participants showed that casuals’ prompts averaged 35 words, while professionals’ averaged 83 words (almost 2.4 times more). This gap reflected not only greater structural detail but also professionals’ use of domain-specific vocabulary and explicit design constraints.



\subsubsection{Seeking Guidance for Prompt Formulation}
The remaining eight participants (7C, 1P) adopted a help-first approach, seeking guidance for input formulation before attempting to prompt the tool directly. Their strategies centered around three resources: external LLMs (ChatGPT), the tool's official documentation on the web, and in-application help features.

\textbf{Use of ChatGPT:} A subset of participants (3C, 1P) engaged in what we term \textit{AI chaining} - the use of an LLM such as ChatGPT to generate or refine prompts for another prompt-based generative tool. Some saw ChatGPT as a faster, more articulate way to produce inputs for generative 3D tasks, \textit{''ChatGPT can describe it faster and better than I can… in a shorter amount of time''}(C03), while others used only after struggling with self-formulated prompts: \textit{''I try [to] create the prompt on my own. And if it doesn’t work… I go to ChatGPT''}(C12). Another participant acknowledged this difficulty directly: \textit{''I will try [to prompt] but it’s a little challeng[ing] to describe what those structure[s are]''}(C13).


\textbf{Official Documentation:} Four participants (4C) began by consulting the tool documentation to learn input format, feature availability, or relevant vocabulary. For some, this was a preparatory step before prompting, but most found the materials unhelpful or inadequate. As C08 remarked: \textit{``not descriptive enough… I have something in my head, but I can’t find it over here''}. Similarly, C02 abandoned the documentation after failing to find usable guidance. Two participants (2C)  shifted quickly to self-prompting, while two casuals (C02, C08) engaged in multi-help sequences, combining documentation and ChatGPT. In both cases, documentation shaped expectations, but ChatGPT supplied the actual text submitted to the tool.

\textbf{Built-in Help: }Notably, no participants used the in-application help features. For example, in Meshy, none of the participants noticed or engaged with the hints icon in the UI (see Figure ~\ref{fig:meshy_interface}). Instead, participants either relied on ChatGPT or their own self-directed experimentation. This pattern aligns with the \textit{active user paradox} ~\cite{carroll}, in which software users, eager to begin their task, bypass available guidance even when it could improve performance. We revisit this point in the Discussion to consider its implications for the design of generative tools.


\subsection{Differences in Trial and Error Strategies for Iteration}

After their first attempt, participants moved into cycles of trial and error, but their strategies diverged sharply by expertise. Casual users approached iteration as open-ended exploration, tweaking small descriptive details, layering on adjectives, or switching between resources such as Google search and documentation when results were unsatisfactory. Professionals, in contrast, pursued fewer targeted refinements grounded in prior CAD and 3D modeling knowledge. They refined prompts around specific parameters such as dimensions, geometry, or material constraints, emphasizing efficiency and structural precision. 



\subsubsection{Direct In-Application Iteration and refinements vs AI Chaining} 
As described above, \textit{AI chaining} was commonly used by a subset of participants before submitting prompts in the 3D modeling tool. In Task 1, 4 participants (3C, 1P) described the approach as a ``faster'' way to ``get the phrasing right'' without spending excessive credits on trial generations. 
Only 1 casual participant used ChatGPT in Task 2, and three participants(3C, 1P) employed it in Task 3.




In contrast, professionals largely avoided external AI support, expressing either distrust in its effectiveness, calling it "unnecessary," or frustration with the quality  \textit{``ChatGPT sucks for 3D design''} (P21, professional). This reflected a broader sentiment that language-model-driven assistance lacked the precision and domain specificity required for complex 3D tasks: \textit{``It's easier for me to put and describe things with words because I had something in mind''} (P16). This preference for self-reliance mirrors established patterns in expert creative practice, where efficiency, control, and trust in one’s own skills outweigh perceived benefits of automated support. 

 When participants did iterate, they pursued two main strategies: (1) \textit{prompt modification} - making incremental changes to the original text; and (2) \textit{prompt restarting} - abandoning the first attempt and writing a completely new description. 

In Task 1, many participants (7C, 7P) wrote the second prompt without relying on external resources, often adding specific descriptors or reordering attributes to prioritize certain features. Casual users typically added visual adjectives in the hope of nudging the AI toward aesthetic fidelity while professionals specified structural constraints table~\ref{tab:prompts_table}. 


The choice between modification and restart reflected how failures were diagnosed. Casuals typically perceived flaws as ``stylistic'' and were more likely to edit; professionals  identified structural inaccuracies or mesh errors discarded the iteration itself. This difference in iteration pathways reflects contrasting expectations: for casuals, it was a creative steering exercise, whereas for professionals, it was a corrective precision task, and in many cases, not worth pursuing if the underlying model quality was judged insufficient:  
\textit{``all of these AI generation tools...generate everything from...scratch. They are using our first prompt...in some cases, we just want to have some changes in some part of the model.(P18)''} Here, iteration was framed less as exploration and more as correction.

\begin{table*}[!h]
\centering
\begin{tabular}{|l|c|c|c|c|}
\hline
\textbf{Approach} &
\multicolumn{2}{c|}{\textbf{Initial Onboarding Action}} &
\multicolumn{2}{c|}{\textbf{Help-Seeking Strategies}} \\ \hline
 & Casual (\%) & Professional (\%) & Casual (\%) & Professional (\%) \\ \hline
No Help Taken & 50.0\% & 91.7\% & 42.9\% & 64.6\% \\ \hline
ChatGPT & 21.4\% & 8.3\% & 8.9\% & 2.1\% \\ \hline
Tool Documentation & 28.6\% & 0\% & 8.9\% & 2.1\% \\ \hline
Google Search & 0\% & 0\% & 19.6\% & 6.2\% \\ \hline
Tutorial Videos & 0\% & 0\% & 3.6\% & 2.1\% \\ \hline
\end{tabular}
\vspace{5pt}
\caption{Comparison of help-seeking approaches across all tasks during the initial onboarding actions and subsequent trial-and-error iterations. Casual users often relied on external AI support (e.g., ChatGPT) and some consulted documentation; professionals mostly began with self-formulated prompts and iterated independently. Unlike prior HCI findings (e.g., \cite{kiani2019beyond}) where video tutorials were a dominant onboarding resource, our participants rarely turned to them. Instead, prompt-based onboarding itself emerged as a central help-seeking strategy, with users treating iterative prompt refinement as both a way to explore system capabilities and as a substitute for formal tutorials. }
\label{tab:help_strategies}
\end{table*}

\subsubsection{Credit-Bound Behaviour and Iteration Risk} Credit-based generation limits shaped the trial-and-error strategies of multiple participants, particularly casual users. Although the modeling tools provided a finite number of free generations, several participants treated credits as scarce resources, conserving them for ``high-confidence'' prompts rather than exploratory adjustments. Some stopped after a single attempt, even when results were imperfect: \textit{``I'm just trying to see like what type of prompts people use for text to image prompts...I don't wanna waste my credits''}(C11); \textit{“If I click it, will it eat up more credits?”} (C02). This often led to premature task closure, with participants accepting suboptimal outputs to preserve credits for later tasks. 
This behavioral shift was also reflected quantitatively in participants’ iteration patterns. In Task 1, participants who were not under credit constraints averaged 4.4 prompt iterations per task, compared to only 2.3 iterations among those who were credit-bound. This suggests that exploration decreased noticeably once participants became aware of credit risk. This pattern intensified in later tasks: by Task 2, 12 participants (6C, 6P) made only a single attempt, and in Task 3, 13 participants (9C, 4P) demonstrated the same one-attempt behavior.


Professionals were also credit-aware, but their reluctance stem-med from doubts about the tool’s ceiling of quality rather than credit scarcity. As P16 explained: \textit{``I don’t think I’m gonna use them [credits], even if they’re free, because my time also has value… if I’m not getting anything even close to what I’m desiring, there’s no point.''} For both groups, iteration decisions reflected resource-rational tradeoffs, where expectations about tool limits, rather than actual capacity, shaped willingness to invest effort.

Rather than empirically testing the system’s potential, participants acted on their mental models of what the tool could or could not deliver: casual users assumed that additional credits were unlikely to be ``worth it,'' while professionals assumed that the AI’s output ceiling was inherently too low. This aligns with broader cognitive theories of resource-rational decision making and satisfaction, where users conserve effort (or credits) once their expectations about system limits are set, even if those expectations may underestimate the tool’s actual capabilities \cite{Resource-rationa, mentalmodel}.

\begin{figure*}[t!]
    \centering
    \begin{subfigure}{\textwidth}
        \centering
        \includegraphics[width=0.8\textwidth]{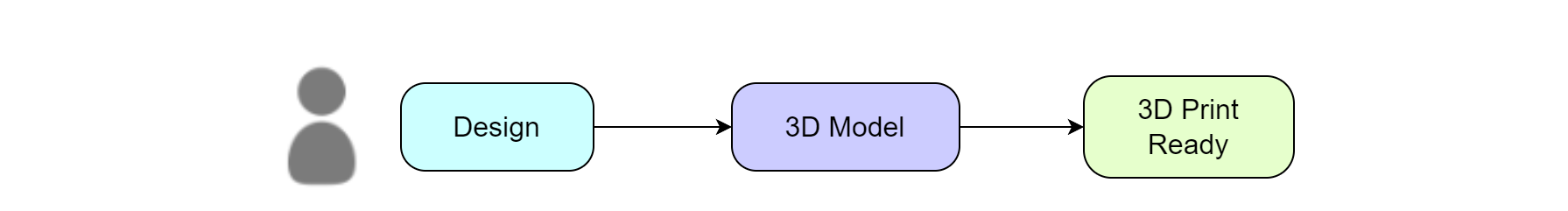}
        \caption{Generalized Workflow of the Casual Participants}
    \end{subfigure}%

     \vspace{5pt}
    
    \centering
    \begin{subfigure}{\textwidth}
        \centering
        \includegraphics[width=0.8\textwidth]{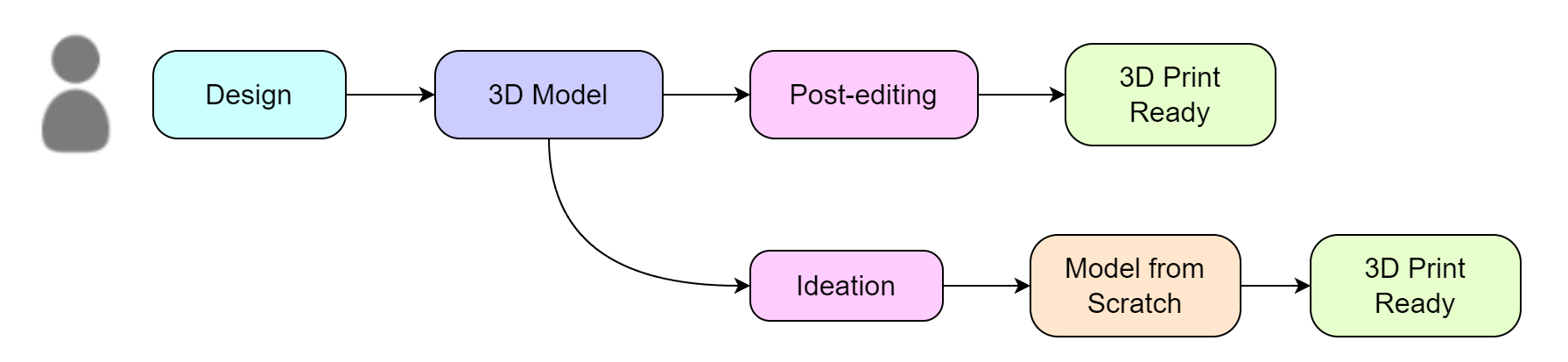}
        \caption{Generalized Workflow of the Professional Participants}
    \end{subfigure}
    \caption{Generalized workflows of participants. (a) Casual participants often treated generated models as immediately ready for 3D printing, overlooking issues such as surface anomalies or structural stability. (b) Professional participants followed two distinct pathways: a post-editing pathway, where they repaired or refined AI-generated models before considering them print-ready, and a model-from-scratch pathway, where they rejected the AI output, returned to ideation, and created new models for downstream use. These contrasting workflows illustrate how casual users emphasized immediacy and usability, while professionals foregrounded production quality and long-term integration.}
\end{figure*}

\subsubsection{Help-Seeking Patterns} 
External help-seeking was surprisingly sparse given that participants were encouraged to use multiple support channels, including built-in documentation, online tutorials, and general web resources (see Table ~\ref{tab:help_strategies}). Google searches were used by six casuals and three professionals across all tasks, mostly to identify terminology (e.g., “vertices,” “cross brace dimensions”). YouTube was even rarer (two casuals total), with participants noting that videos were too slow for time-limited tasks. Official documentation was also underused, accessed by only four participants in Task 1 and fewer thereafter, often via Google rather than the tool’s own interface. Participants made comments such as, \textit{``if I need something urgently, I don't have time to go and learn the resources''} (C08).
This number dropped to 3 casuals in Task 2 and 2 participants (1 casual, 1 professional) in Task 3. A common complaint was the unfamiliarity with the jargon:  \textit{``I have something in my head, but I can't find it over here [documentation] I don't know what is what.'' (P08)} Another participant expressed the hesitation to use tutorials: \textit{``I feel like that the tutorial section was more about... how to make certain things, but I kind of wanted...more general bit, like how do I use the different aspects of the AI tool''} (C20).  

Casual users were more inclined to experiment with layered help-seeking, often moving between multiple resources within the same task. For example, in Task 2, C02 sequentially consulted the Spline documentation, then searched Google, moved to YouTube, and finally used ChatGPT to refine their prompt after repeated mismatches with the target model.  C12's narrative captures the sentiment common among casuals: 
\begin{quote}
    \textit{``Youtube tutorials are helpful when you're a novice, and you just want a general walk through, an orientation and have someone explain everything to you. But also AI tools like Chatgpt are really helpful if you're stuck on a specific scenario, or I'm trying to do this exact thing in my project, and I don't know where to click this or that it can actually give you pretty good step-by-step instructions''}
\end{quote}

In contrast, professionals showed minimal reliance on external resources (only 3 professionals used Google at any point). These users expressed confidence in their own ability to diagnose and address modeling issues, even when confronted with unsatisfactory AI outputs. Several described external resources as \textit{too slow } or \textit{not specific enough} for their precision-driven needs, preferring instead to either restart from scratch or revert to traditional CAD workflows.

\subsection{Differences in Workflow Perceptions}

Participants’ interpretations of AI-generated outputs revealed a clear divide in how casual and professional users: (1) evaluated AI outputs, and, (2) where they positioned the tool within their broader workflows.

\subsubsection{Evaluation Standards: Good Enough vs. Production-Ready} 

Casual participants (n = 14) often treated generated outputs as finished artifacts, suitable for immediate 3D printing or display. For them, visual resemblance was the dominant quality metric: if the model “looked like” the intended object, it was assumed to be fabrication-ready, regardless of mesh integrity, wall thickness, or structural stability. As C12’s remarked: \textit{“I was relieved at how easy it was to create an object just with AI prompts… usually you have to get really familiar with tools, learn how to create stuff from scratch.”} Similarly, C08 found that getting an "approximate" match was sufficient: \textit{“It was able to get the basic idea in check… even though it might not be 100\% similar, it was good enough so I could convey my idea.”} 

Professionals (n = 12), in contrast, consistently judged outputs as not production-ready in their unedited form. Professionals' assessments focused on structural quality and fabrication feasibility, systematically inspecting geometry for errors (e.g., non-manifold meshes, thin supports, or irregular topologies). P21 was blunt: \textit{“This is not usable in a final product at all.”} P17 highlighted issues of printability: \textit{“These smaller details around here? It might make 3D printing hard.”} P06 emphasized watertight geometry: \textit{“To have anything 3D print ready… there should be no thinning or holes. Like, it should be a solid model. With this, I don’t think it is.”}  

\subsubsection{Workflow Positioning: Direct Output vs. Early-Stage Ideation} 

The differing evaluations of the output of our participants shaped distinct workflow trajectories. Casuals often pursued a minimal sequence of prompt → print, skipping intermediate steps such as inspection or refinement. As C08 explained, speed outweighed precision: \textit{``I think for simpler things, like… a cupcake… this is good. It’s available widely. I can do it very fast.``} The ability to bypass labor-intensive modeling became the main source of satisfaction, redefining what “\textit{good enough}” meant in their workflows. 

Professionals instead positioned the AI tool at the ideation stage of their workflow, where the outputs could serve as inspiration. Three participants (P09, P17 and P18) attempted post-editing (Design → 3D Model → Post-Editing → 3D Print), exporting models into advanced 3D modeling tools, such as Blender, SolidWorks, or Rhino, for hole filling, edge merging, and wall-thickness adjustments. However, this was time-consuming and prone to new errors (e.g., corrupted UV maps). The more common pattern (9 out of 12  professionals) was to abandon editing entirely, following Design → 3D Model → Ideation → Model from Scratch → 3D Print. As P17 noted: \textit{``If it's like bumpy, and I have to clean it up myself, that would be kind of annoying.``} P21 similarly emphasized that AI benefits only at the earliest stages: \textit{`` only saves me time in the fastest ideation portions.``} 

This distinction echoes prior work on casual makers in walk-up-and-use 3D printing environments ~\cite{hudson2016understanding}, where casuals often avoided refinement altogether. In contrast, for professionals, AI outputs offered conceptual value but rarely fabrication utility. In both cases, workflow breakdowns curtailed deeper design exploration, casuals due to usability hurdles, and professionals due to tool-chain complexity. 


\renewcommand{\arraystretch}{1.2}
\begin{table*}[ht]
\centering
\caption{Key Insights of casual and professional participants engaged with generative 3D modeling tools. The comparison highlights distinct patterns across initial engagement, prompt style, help-seeking behavior, iteration strategies, workflow integration, and output evaluation, reflecting the contrasting approaches and expectations shaped by varying levels of expertise.}
\label{tab:key insights}
\begin{tabular}{p{0.18\textwidth} p{0.39\textwidth} p{0.39\textwidth}}
\toprule
\textbf{Dimension} & \textbf{Casuals} & \textbf{Professionals} \\
\midrule
Initial Engagement & Begin with a quick, self-formulated prompt or seek help from ChatGPT or documentation before starting. & Usually start with a prompt-first approach, relying on their own descriptive skills without consulting help. \\
\midrule
Prompt Style & Prompts tend to be shorter, less structured, and more visually descriptive. & Prompts are highly detailed, structurally explicit, and specify precise geometric or dimensional constraints. \\
\midrule
Help-Seeking & Frequently use layered help sources like ChatGPT, Google, and documentation, often in ad-hoc sequences. & Rarely seek external help, preferring self-reliance and domain expertise over outside resources. \\
\midrule
Prompt Iteration & More likely to make small wording tweaks or stylistic edits, sometimes chaining AI tools for refinement. & Prefer major rewrites or start from scratch when outputs deviate structurally, avoiding external AI tools. \\
\midrule
Workflow Integration & Treat AI-generated outputs as production-ready, moving directly from generation to intended use. & Use outputs as early prototypes or concept references, requiring significant post-editing or full remodeling. \\
\midrule
Output Evaluation & Judge quality mainly by visual similarity to the intended object, often overlooking structural flaws. & Assess outputs against strict manufacturability standards, identifying geometric and structural issues before use. \\
\bottomrule
\end{tabular}
\end{table*}

\section{Discussion}

\subsection{Larger Context of Our Findings}

Our study reveals a notable shift in how users begin learning and seeking help with creative software: rather than relying on videos, forums, or search engines as documented in prior HCI work \cite{kiani2019beyond, mahmud2020learning, tovigrossman}, participants increasingly turned to other generative AI systems to understand or operate the primary 3D tool. Casual participants, in particular, rarely explored Meshy or Spline through their interfaces; instead, they moved directly into AI chaining workflows for instance, asking ChatGPT to rewrite prompts, fix prompt phrasing, or explain unexpected model outputs before returning to the primary 3D tool. Professionals engaged in less external AI use but still relied on in-application AI features for prompting and making trial and error rather than documentation or tutorials. Taken together, these patterns suggest a key shift in onboarding from feature exploration \cite{predictivemenu, sixlearning, Lafreniere} toward prompt formulation as the primary entry point. As more modeling platforms integrate embedded copilots \cite{ barke2022groundedcopilotprogrammersinteract}, model-assisted refinement \cite{liu20233dalle}, or auto-complete designs \cite{chen2016reprise}, we expect this GenAI-to-GenAI help ecosystem to become an increasingly dominant mode of onboarding and problem-solving.

Despite these shifts, our findings also show continuity with long-standing theories of how users learn and interact with feature-rich software. Participants skipped tutorials and avoided in-built tool documentation in favor of immediate action, consistent with minimalist instruction \cite{carroll1998minimalism} and the paradox of the active user \cite{carroll}, which predicts that users prioritize “getting started” over learning even when it leads to inefficiency. This preference for immediate action was reinforced by production bias, the tendency to favor quick progress over deeper understanding and structured learning \cite{carroll}, which was especially visible among casual users who aimed to obtain a usable model as quickly as possible, even if that meant relying on imperfect prompts or overlooking system features.  Likewise, users’ help-seeking patterns reflect well-documented self-regulated learning processes  \cite{nelson, Aleven}: users monitored breakdowns, evaluated failed outputs, and selectively turned to external assistance, often alternating between instrumental help-seeking (seeking understanding, more common in professionals) and executive help-seeking (seeking fast solutions, common among casuals). Taken together, our findings suggest that while generative AI introduces new mechanisms such as AI chaining and language-based onboarding, the underlying motivations that shape users’ software learning and help-seeking behavior remain well-accounted for by earlier conceptual and theoretical observations in HCI.

\subsection{Help-Seeking as Multi-AI Ecosystem Behavior}


Our findings suggest a reconfiguration of help-seeking for feature-rich software, one defined not by consulting manuals or tutorials but by coordinating assistance across multiple AI systems \cite{kaptelinin2006acting}. Rather than relying on a single source of support, participants treated different generative tools as complementary resources, each offering partial guidance or corrective feedback.  AI Chaining was not a workaround but an intentional strategy shaped by credit limits, uncertainty in outputs, and the opacity of individual models. These patterns suggest that HCI should increasingly understand help-seeking as a distributed process and treat AI tools not as standalone systems but as interdependent collaborators within user-constructed help ecosystems.

\subsubsection{ \textbf{Design Implication:} Designing Workflow-Level Multi-Model Guidance}
Our findings suggest that Gen AI  tools should no longer treat the creation pipeline as a single opaque step, but instead surface the underlying workflow in a way that supports both learning and expert control. By integrating this behavior into the interface, tools could make each stage of the generative pipeline visible and optionally editable, allowing novices to follow a guided, progressive path (generate → refine → texture → export) while enabling experts to collapse, bypass, or reorder stages. This approach could build upon established principles of progressive disclosure and staged workflows in complex software \cite{tovigrossman, findlater} , as well as distributed cognition perspectives showing that users naturally offload cognitive work across multiple artifacts when solving complex tasks \cite{Hutchins}. Rather than hiding intermediate steps, systems could offer concise explanations of what each model is doing, show intermediate artifacts, and provide diagnostic feedback at transition points helping users build accurate mental models of the process. Such workflow-level guidance also creates a foundation for explicit handoffs between AI tools, acknowledging the reality of multi-model use and reducing the trial-and-error burden users currently face. By making the generative pipeline transparent and adaptable, tools can better support the very different expectations casual and professional users bring to AI-assisted 3D modeling.

\subsection{ Rethinking Software Onboarding in the Age of Prompts}

Instead of engaging with tutorials, interface walkthroughs, or feature exploration, we found that users begin by producing language instructions. This pattern shows a shift in onboarding from learning how to operate a tool to learning how to express intentions in a way the AI can interpret. This reframes onboarding as a negotiation with the language–action gap of generative models rather than the interface itself \cite{PEREIRA2023100857, Khurana_2024}.  For casual users, this shift lowers the barrier to `getting started,' because the entry point becomes conversational rather than procedural. For professionals, however, this inversion complicates expectations of precision, reproducibility, and control qualities traditionally grounded in direct manipulation workflows. Professionals’ expertise, built around spatial reasoning, constraint management, and iterative refinement, does not always transfer into a text-based interaction model. 
Together, these patterns illustrate a new form of software learning: one where the primary challenge is not feature discovery but understanding how to collaborate with a stochastic system whose internal logic is opaque. This creates a divergence in experience: beginners benefit from reduced complexity, while experts encounter friction when the model fails to match their expectations for accuracy, parameter control, or workflow alignment.

\subsubsection{\textbf{Design Implication:} Designing Micro-Scaffolds That Leverage the Prompt Loop}

 Traditional contextual help has existed since early intelligent tutoring systems \cite{Aleven}, but generative AI creates an opportunity to design in-time micro-scaffolds that work within the user’s existing prompt-iteration workflow rather than interrupting it. Instead of surfacing generic hints or feature explanations, systems could provide prompt-level scaffolding that adapts to the user’s pattern of failures. For instance, they could surface example prompts when the model detects repeated semantic misunderstandings, provide parameter suggestions when outputs show consistent geometric issues, or highlight generative constraints (`thin surfaces', `unsupported geometry') as lightweight annotations on the prompt itself. This idea could further build on research showing that productive help is most effective when it is embedded in the user’s immediate task state \cite{Aleven} and when it supports users’ existing strategies rather than requiring entirely new ones. By designing scaffolds that amplify the user’s natural “learn-by-prompting” loop rather than attempting to reintroduce traditional tutorial models, tools can better support both casual users who rely on rapid experimentation and professionals who require conceptual clarity to maintain precision. In this sense, the implication is not simply to “add contextual help,” but to rethink help as prompt-guidance infrastructure, a direction aligned with adaptive interfaces \cite{findlater} and modern HCI research that treats AI as both collaborator and instructor \cite{JIANG2024100078}. 



\subsection{`Good Enough' vs `Not Even Close': Divergent Standards of Sufficiency of AI Output
} 
Our findings reveal a clear divergence in how casual and professional users evaluate the sufficiency of generative 3D outputs. Casual users frequently accepted visually plausible models as “good enough,” collapsing the traditional design pipeline into a direct path from concept to artifact. In contrast, professionals judged the same results as structurally inadequate or unsuitable for downstream workflows, identifying issues such as non-manifold geometry, missing supports, or inaccuracies that compromise printability. This division echoes prior observations in computer graphics and vision that generative models optimize for visual plausibility rather than geometric correctness \cite{sitzmann2020scenerepresentationnetworkscontinuous}, and that mesh-quality issues remain a persistent challenge even in state-of-the-art 3D generation pipelines \cite{zhou2023instructionfollowingevaluationlargelanguage}\cite{jun2023shapegeneratingconditional3d}. By examining user workflows and help-seeking patterns, our study shows that “visual realism” valued in generative AI research does not translate uniformly to “workflow-ready quality” valued by professionals. This underscores why human-centered evaluation is essential: while generative 3D lowers barriers for newcomers, it also widens the experiential gap between novices and experts by privileging immediacy over precision. Designing systems that support both groups within the same interface remains an open challenge, requiring tools that recognize divergent standards of sufficiency rather than treating output quality as a single, universal metric.

\subsubsection{\textbf{Design Implication:} Designing Dual Modes for Concept vs Production Workflows}
Our findings suggest that generative 3D modeling tools would benefit from explicitly supporting two parallel modes: one optimized for rapid, visually plausible concept generation and another designed for structurally reliable, production-ready modeling. Introducing a Concept Mode would embrace these strengths by optimizing for immediacy, creative exploration, and low cognitive overhead, echoing prior HCI work showing that novices benefit from simplified workflows, reduced parameter exposure, and rapid feedback  \cite{findlater, tovigrossman}. In contrast, a Production Mode could surface the kinds of controls and diagnostics that experts expect in precision-oriented environments, such as mesh manifold checks, thickness warnings, printability indicators, and parametric constraints, aligning with practices from CAD, animation, and fabrication pipelines where structural integrity and export fidelity are critical \cite {attene}. Supporting these two modes allows systems to respect the divergent standards of sufficiency of AI output among casual and professional users. Prior HCI studies on user interaction with feature-rich software show that when interfaces cater exclusively to novices, experts experience friction and inefficiency \cite{kiani2019beyond, Fitzmaurice}. In contrast, enabling users to fluidly switch between conceptual and production workflows acknowledges both ends of the expertise spectrum without forcing a one-size-fits-all interaction model. Such dual-track design could also offer a more meaningful interpretation of “democratization,” ensuring that generative tools empower newcomers without compromising the rigor necessary for professional practice.

\subsection{Credit Constraints and the Dynamics of Generative Exploration} 
We found that credit-based constraints significantly shaped how users explored generative 3D tools, especially among casual users who became risk-averse and limited iteration to avoid “wasting” credits. Although our study focused on 3D modeling, this pattern could generalize to other generative domains where outputs are token-metered or computationally expensive (e.g., AI-assisted CAD, texture generation, and parametric design). Our study suggests that credit constraints amplify production bias, heighten loss aversion, and shape trust decisions, pushing users to judge whether an output is “worth a credit” based on surface plausibility rather than structural correctness. Nevertheless, domain-specific constraints may influence how strongly these behaviors appear.

\subsubsection{\textbf{Design Implication:} Designing for Credit Transparency and Expectation Calibration
}
To counter credit-driven constraints, generative tools could make credit consumption more transparent by previewing the expected credit cost of an action, the likely fidelity or structural reliability of the output, and lower-cost alternatives before generation occurs. Such predictive scaffolding echoes principles from trust-calibration literature, which emphasize that users form more resilient trust when systems communicate uncertainty, limitations, and probable failure modes upfront rather than retroactively \cite{norman2013doet}. For example, when a user attempts to generate a mechanically precise or physically constrained object, the system might warn that generative models typically struggle with thickness tolerances, manifolds, or print-ready geometry, issues well-documented in 3D generative modeling research \cite{jun2023shapegeneratingconditional3d, zhou2023instructionfollowingevaluationlargelanguage}. By framing credit cost and output reliability together, systems can help users make informed trade-offs between exploration and resource consumption. This is particularly important because, as our findings showed, uncertainty about credit consequences caused participants to prematurely settle for suboptimal outputs or rely on external AIs (e.g., ChatGPT) as a no-cost workaround. Transparent cost-benefit signals, therefore, not only reduce anxiety around credit spending but also support more deliberate exploration, encourage productive iteration, and calibrate user expectations toward what generative 3D models can and cannot reliably produce. Designing for such calibrated expectations is essential for building trust without fostering overconfidence, a challenge repeatedly highlighted in recent work \cite{Kahr, lai} on AI-assisted creativity and decision-making.

\subsection{Limitations}
Our study has several limitations that frame the scope of our findings. First, while the participant pool included both casual and professional 3D designers, our professional sample covered a range of roles such as product design, CAD engineering, and 3D artistry, but not all subdomains such as game asset modeling or large-scale fabrication. Second, the study focused on two specific generative 3D modeling tools. Although these platforms are representative of current systems, differences in algorithmic back-ends, mesh quality, and interface design could yield different user behaviors. Third, our task-focused study design may have influenced iteration behaviors. In real-world practice, users often work under time, budget, and collaborative constraints, which could amplify or dampen the strategies observed. For example, professionals’ may view post-editing in a commercial context where AI outputs could serve as rapid prototyping starting points under strict deadlines. Finally, our credit-based conditions reflected the default limits of the tools rather than an experimental manipulation; future work could vary credit availability more systematically to assess its impact on exploration and help-seeking. Additionally, the fixed task order may have introduced order effects. We prioritized a consistent progression across tasks to ensure comparability; however, this may have unintentionally amplified prompt-first behaviors early in the session and suppressed later exploration or help-seeking due to time pressure or credit depletion. As a result, some participants may have relied more heavily on initial prompting strategies rather than revisiting tutorials or alternative workflows later. Future work could counterbalance task order or vary credit availability to better isolate these effects.
We made a proactive effort to recruit participants across genders and age groups, though most ultimately reflected the younger demographic that currently dominates early adoption of generative 3D modeling tools. Some groups remain underrepresented due to the emerging nature of these systems and the characteristics of their present-day user base. As generative 3D tools mature and attract broader audiences, future work should expand demographic diversity to capture a wider range of perspectives.

\section{Conclusions}
Our study shows how generative 3D modeling reshapes design by collapsing the boundaries between learning, creating, and seeking help. These shifts are less about skill than about different visions of AI as a shortcut, collaborator, or unreliable partner. Participants often orchestrated multiple AI systems in tandem, pointing to a future where help-seeking extends beyond manuals or forums into distributed dialogues across AI tools. For designers of creative technologies, this means building systems that not only generate results but also adapt to diverse mindsets, supporting both users seeking ready-made outputs and those treating AI as a design partner. For HCI, the broader challenge is to see AI ecosystems as emerging sites of creative practice where expertise, authorship, and support are being redefined. Attending to these transformations is vital if HCI is to keep pace with how generative AI reshapes everyday design work. Ultimately, design is no longer solitary; it happens in concert with AI.

\section{Acknowledgments}
We thank the Natural Sciences and Engineering Research Council of Canada (NSERC).

\bibliographystyle{ACM-Reference-Format}
\bibliography{bibliography}

\appendix









\end{document}